\documentclass[preprint]{aastex}

\usepackage{verbatim}

\usepackage{listings}
\usepackage{color}

\usepackage[yyyymmdd,hhmmss]{datetime}

\shorttitle{A New Framework for Doing Science With Data / Computational Platform}
\shortauthors{D. Muna $^{1}$ and E. Huff$^{1}$}

\begin{document}

\title{A New Framework for a Model-Based Data Science Computational Platform}

\author{D. Muna}
\affil{Department of Astronomy, Ohio State University, Columbus, OH}
\and

\author{E. Huff}
\affil{Center for Cosmology and Astroparticle Physics, Ohio State University, Columbus, OH}


\begin{abstract}
Astronomy produces extremely large data sets from ground-based telescopes, space missions, and simulation. The volume and complexity of these rich data sets require new approaches and advanced tools to understand the information contained therein. No one can load this data on their own computer, most cannot even keep it at their institution, and worse, no platform exists that allows one to evaluate their models across the whole of the data. Simply having an extremely large volume of data available in one place is not sufficient; one must be able to make valid, rigorous, scientific comparisons across very different data sets from very different instrumentation. We propose a framework to directly address this which has the following components: a model-based computational platform, streamlined access to large volumes of data, and an educational and social platform for both researchers and the public.
\end{abstract}



\section{Introduction}

The era of big data is upon us. Existing astronomical surveys like the Sloan Digital Sky Survey (SDSS), with a data volume of $\sim$15 TB, already stretch the limits of what our analysis methods and software are capable of. Next-generation surveys, such as the Large Synoptic Survey Telescope (LSST), will generate an SDSS every night \citep{2008arXiv0805.2366I}. How is a research astronomer to cope? The mishmash of mismatched survey outputs from data sets documented to varying degrees with differing catalog conventions, clunky web forms, small departmental computing clusters, and roll-your-own analysis software already wastes enormous quantities of time that could be better spent on actual science: building and testing models, looking for unexplained features, and using data visualization to build physical intuition and motivate new understanding.

We present here a new framework that addresses the major challenges of doing astronomy with big data. The framework can be roughly described as having three primary components: a model-based computational platform, streamlined access to large volumes of data, and an educational and social platform for both researchers and the public. It will let theorists compare physical models to data from a wide variety of astronomical surveys without having to understand for each survey the point-spread functions, photometric calibration, or the effect of aperture size choice on photometry. It will allow researchers to write code to run on data as if it were local, without the hassle of managing hundreds of terabytes of local data volume. Our framework will attach quantitative models and their associated fit likelihoods to astronomical objects, letting researchers ask questions like ``How many stars have colors or spectra that look like my model?''~or ``What things in the universe are unexplained by existing models?''~with straightforward database queries.  Associating quantitative models, their constituent code, and their detailed descriptions with these objects creates a mechanism to search for and compare with the relevant literature, and it provides a path into the subject for educators and the general public.

We will start by identifying current problems in data analysis, offer a high-level overview of the proposed framework, then describe the implementation of the framework in greater detail. Astronomical data sets and models are used as examples to present the ideas in a more concrete manner as this is the primary domain of the authors, but nothing presented here need be specific to astronomy. This framework as offered can be extended such that the data sets and models of many fields may overlap in new and novel ways. While pieces of this framework, such as designing large databases, data visualization, etc.~are certainly being developed in fields such as astrophysics, computer science, and human interface design, the approach described here fully integrates these and other ideas to enable new and novel scientific research and education.

\section{Workflows: Today and Tomorrow}
\label{section:workflows}

Rather than begin in the abstract, we start by describing a typical workflow of an astronomer using tools available today. Following this is a description of the same task performed using the framework proposed here. The rest of this paper will detail the motivations behind the framework, its aims, a detailed description of its implementation and features, and a discussion of new questions and considerations it carries with it.

Astronomer Anne Elk is a theorist interested in locating a population of stars that are one billion years old. She generates a theoretical evolutionary history of a type of star, which produces parameters, e.g.~temperature, surface gravity, luminosity, in natural units. Using these values, Anne creates a set of stellar spectra at a fixed age of 1By. She then must decide which catalog to use for the search. Choosing SDSS, she converts the colors into the \textit{ugriz} filters specific to that instrument, a step that requires specialized knowledge of the survey. With these values, she goes to a web interface to create an SQL query for objects that match these magnitudes. Anne prefers to use an extinction method that employs magnitude values from NASA's Wide-Field Infrared Survey Explorer (WISE) satellite and the Two Micron All Sky Survey (2MASS), a calculation that must be done on a star by star basis. This involves looking up these values from two more web interfaces for each star. Since this correction can only be applied after the search in the SDSS data, she expands the search allowing for a maximum extinction correction. She downloads the result of the SDSS query, but since the initial query exceeded the limit available to users, she must repeat these steps until she can download all of the potential targets. Next, she creates a new file of the coordinates found and uploads this target list to another web form to retrieve the WISE magnitudes, and repeats this for the 2MASS magnitudes. Most of the targets match a single object in these catalogs, but some match a few and must be handled by hand. Again, the list is too long, and she must repeat these steps to get the full target list. Once she has downloaded this information, she must then write code to calculate the extinction for all of the candidate stars, then filter those that fall out of the color range she determined earlier, finally producing a candidate population of 1By old stars.


The goal of the framework we present here is to allow Anne to work within her mental model. Using this framework, Anne would create a representation of her physical model, independent of data -- in this case, a set of spectra for the stars she's interested in, and a flag that she's looking for point sources, though for other projects she might provide a piece of code that represents a parameterized image or spectral model. The files that define this model, the code, and a written description (including literature references) are packaged together into a module and submitted to the framework. This in turn feeds all available catalogs into the model and generates a likelihood for each input. The model is publicly available, so that anyone interested in modifying it, reproducing it, or building new models on top of it can. In fact, Anne's model has as a dependency an existing package in the system that has precalculated extinction values

She opens a client application written to be a front end to the framework to see her results. She selects her model and sees how many objects were evaluated (750 million), and how many CPU hours were used to perform the calculation. She then filters the results to those with a likelihood of greater than 0.85. She plots the results on a color-color diagram; the same application knows how to communicate with the remote server and pulls the values to plot directly, saving the need to download the data to visualize it. Hovering her cursor over one point reveals details about it: the coordinate on the sky, its 2MASS ID, SDSS magnitude, the Hipparcos distance, etc. Clicking an icon fetches an image of the star, allowing her to quickly toggle between Digitized Sky Survey (DSS), SDSS, and WISE images. The program also shows that the star has been compared to several models submitted by other people; one identifies stars on the main sequence, another that identifies brown dwarfs, and a dozen others. Models are listed in order of the likelihood for each model. She clicks on the main sequence model, and finds a detailed description of how that model was designed, as well as a detailed description of what the main sequence is. After looking at several plots, she finds outliers which inform her choices as she refines her model. She emails a note to her colleague and he begins to look at the results on the same system having only been given the name of the model she created.

In the latter scenario, the scientist spends far more time considering her model and data in terms of physical quantities. The framework will know how to work with data in physical units and can determine the magnitudes and colors in each band of whatever data is available. It is designed to handle predictions that are as close to idealized, physical models as possible. The theorist shouldn't have to worry about extinction, photometric calibration, point spread functions, etc. Computing a likelihood requires knowing all these things, but the framework will do the work of mapping the physical prediction onto the predicted image or spectrum, and then compute a likelihood for the data. The details of managing the data --- web forms, downloading subsets of catalogs, writing code to perform cross references, understanding how to directly compare measurements from different instruments each with different point source functions, manually downloading image data, and the tedious bookkeeping involved --- are all performed by the framework. The process of searching for data that matches a particular model is exactly the same for a two degree square field as it is over the entire sky. 

\section{The Problem with the Status Quo}

\subsection{Motivation}

Most scientists have been able to keep copies of data needed for analysis, e.g.~collected or downloaded, on their own computer. Access to data fifteen years ago mainly took the form of downloading flat files, but most people's workflow has not changed significantly (or for many, at all), since then. Even as data sets have grown dramatically larger, many make do by collecting a subset of the available data which enables the continued use of the same analysis tools. However, data sets have now become so large that not only are they greater by orders of magnitude than what's available on personal machines, they are significantly larger than the storage space available to all but the largest university departments. For example, the spectral and imaging data of WISE and the SDSS survey alone exceed 35 terabytes. Even if this information were available locally on disk, the majority of researchers do not know how to query this volume of data. The ability to fully analyze large multi-wavelength data sets thus immediately has a high cost of entry with the expense of computer hardware, management, maintenance, visualization, etc., creating a significant barrier to institutions who do not have that level of resources or scientists who lack the technical skills to create the necessary tools. While we strive in astronomy to make data publicly available, these limitations impede the real motivations behind the ethic, that is, science should be democratic, and open to all. Transparency and reproducibility will only be increased if people who don't have the resources to own systems that cost many thousands of dollars are still able to rerun analyses and experiments.

\noindent We list here the aims that have motivated the data analysis framework presented in this paper:

\begin{enumerate}
\item Enable full access by anyone to the most important publicly available data sets in such a manner that the data appears to be local.
\item Make the amount of work required to analyze any data set the same, regardless of size.
\item Enable researchers to work as closely to the mental model they have of their data (e.g.~physical units or theoretical models) by implementing the specialized knowledge of detectors or instruments in such a way that they can be used by the community.
\item Provide a framework that allows disparate data to be directly compared.
\item Create visualization tools that that are extensible, can scale with large data sets, and provide functionality that is common to and can be shared across all disciplines.
\item Make scientific model contributions more visible and provide a new evolving, educational platform.
\end{enumerate}

\subsection{Data Access}
\label{section:data_access}

Science researchers currently access data in several ways. One is to download some subset of one or more data sets of interest onto their own system, requiring users to manage it themselves, write custom code to read it, perform cross references, etc. This by itself can be challenging as few scientists have functional knowledge of databases, thus the resulting code is often inefficient or doesn't scale to larger data sets. Another model presents the user with a web interface allowing queries of variable complexity. The request is submitted to a queue, the user waits for the job's completion, interactively downloads the data as a file, repeats this if the results exceed the limits of service, joins the results together, writes code to read the files, possibly filter out results as compared to local data, and only then is ready to begin to analyze---or even visualize---the data. Upon analysis, typically a new query is needed, and the entire process is repeated. This method of data access is slow, repetitive, and offloads too much tedious work to the user. Further, since the data subsets downloaded are often simple flat files, they do not contain the steps or query that generated the data, adding another bookkeeping step or worse, the risk of forgetting how the data set was generated. Unfortunately, this scenario is one of the better ones; many data sets don't have even a minimal web interface; without this, users must download large amounts of data and/or write custom code to traverse filesystems or file servers, effectively writing a custom search engine on flat files. This is a waste of time with a large multiplier on the number of users of such data, particularly since search engines are a solved problem.

To some extent, this state of affairs is understandable; the majority of money available to a given survey can and should be maximized to get the most and best data available, and there are few to no resources to provide modern interfaces to and visualizations of that data. Still, these are sorely needed, and common efforts that span multiple data sets are likely the best way to a solution.

\subsection{Linking Theory to Data}
\label{section:linking_theory_to_data}
Local access to large data sets is only the first problem. Astrophysical models predict physical, idealized quantities, such as the temperature and surface gravity of a star or the mass of a dark matter halo. Converting these physical predictions into a vector of pixel values is not trivial, and requires a sophisticated model of the data-generating process. Fitting stellar population synthesis models to multi-band galaxy photometry, for instance, requires accurate models of the telescope point-spread function (PSF) (which will in general be different for each separate observation), the image noise properties, and the sky level. This knowledge is usually generated by the survey, but the scientists with the code to do so are in general not the same people who produce the models in the first place. Simply fitting a theoretical model to observations from multiple surveys requires a significant amount of collaborative effort. There is no reason, in principle, why theorists cannot be given an interface to directly interrogate the data, but at present this is simply not done.

\subsection{System Resources}
\label{section:computation_resources}
The resources required to support the handling of very large data sets (where ``very large'' can be as little as 10 terabytes) are beyond most individual departments. The cost to purchase and maintain servers, disk space, compute clusters, and support staff means that the barrier to entry can be high. Even if these were not issues, to have dozens or more departments maintain redundant copies of large catalogs is a waste of resources. All of the large public data releases are read-only and effectively static, and the larger the data set, the more likely it is that large parts of the data are rarely accessed. Having a complete set of data is valuable, but the rate of access is relatively low. Similarly, unless a department has a theory group running computationally intensive modeling, powerful 16-core servers often spend a lot of time running idle. Instead of a disjoint collection of expensive redundant systems, a system that is shared across the community is far more cost effective, will be much more efficiently utilized, and be more powerful than any one department could afford.

\subsection{Hypothesis Generation \& Interactive Data Visualization}
\label{section:data_viz}
The exploration of data goes well beyond the ability to store or even access it; it must be visualized as well. The ability to ask questions of data and probe the relationships between parameters is a fundamental part of analysis and building hypotheses. It is iterative as well; one plot will lead to more questions, more plots, more questions, etc. Visualization software thus should be as easy to use as possible, allowing and encouraging a flexible, interactive interface. It's important to remove as much code/interface/limitations that lie between the researcher and the data. One aspect of this is the ability to handle large amounts of data; ones that may be greater than the memory on a user's computer, let alone ones that exceed the total storage capacity of their department's servers. Even then, putting several million points on a plot is not likely to reveal a clear picture of the data; even the humble scatter plot must be upgraded with interactivity. New plotting and visualization tools must be developed that can scale from local ASCII files to multi-terabyte data sets that span across institutions. A combination of caching, remote data access, and server-side processing are all interesting avenues that should be explored to satisfy this need.

Consider two modes of scientific research; model fitting and hypothesis generation. The first begins with an existing model which is applied to a set of measurements, for example, fitting a blackbody curve to flux values at different wavelengths or applying an isochrone to a stellar population. Hypothesis generation, on the other hand, is a process of exploration to produce a hypothesis to explain the underlying nature of the data. This is done through the generation of plots, but more generally, it is a process of reducing the complexity and dimensionality of the data to match that of a developing mental model. For example, a data set of stars containing temperature, luminosity, surface gravity, age, mass, composition, and metallicity has a high dimensionality; it is too complex to visualize the ensemble all at once. Instead, one looks for relationships between parameters in pairs to explain the underlying physics. Existing tools require far more manual work than should be necessary, and simple but fantastically useful techniques like data brushing, while not new, are still unheard of by a large number of scientists.

While it is true that no one plotting program can satisfy the needs of any arbitrary analysis, there is a large domain of visualization that is common across all sciences. Given a data set with a high number of dimensions, scatter plots, histograms, model fitting, data brushing, etc.~are all tools that are not domain-specific. Commercial solutions do exist but are prohibitively expensive (assuming they are available on your platform), particularly in an environment where everyone is expected to write their own code. It is surprising---or at least should be!---that there are no sophisticated, community, open source data analysis plotting programs. Plots can be created with software such as R and Python's matplotlib, but they require each person to code their own visualizations essentially from scratch. While these languages have the ability to connect to external data sets from a database or over the internet, programming techniques such as these are rarely taught.

We as scientists should develop and make such tools easily extensible, and widely and freely available. Approaching this from a multidisciplinary point of view will enable new and novel ways to visualize data that might not already be used in a particular field. It's not an understatement to say that such a tool would save an enormous amount of time as a multiplier against thousands of scientists; whatever the cost of development, it will pay for itself in productivity many times over.

Finally, to address the astronomical community specifically, we feel that visualization of data written in the Flexible Image Transport System (FITS) format is long overdue for a complete overhaul. This format is the workhorse of nearly all astronomers, yet the most common tools are written in X11, an obsolete framework that is thirty years old. Many astronomers dislike the FITS format, but we contend that the tools are a much greater concern than how the bytes are stored on disk. While there are good C and Python interfaces to FITS (which are extremely valuable), these are programmatic in nature, requiring code to be written and run to do even the simplest tasks. The community should support the development of new software, written from scratch, where each and every interface design choice is reevaluated rather than copied from the past. The software should be able to generate plots, display statistics of the data found in the file, communicate with remote servers to cross reference data from point source catalog and image servers; in short, the tasks common to most anyone who uses such files.

\section{A New Data Science Model for Research}
\label{section:solution}
\subsection{High Level Description of Deep Thought}
We propose a new framework that will address the problems in research highlighted above, make some types of existing data analysis considerably easier and more powerful, and provide a means to query large data sets in a way that is either extremely difficult or impossible now. When people think of a scientific model, their mental representation is in physical terms inherent to that model, e.g. temperature, element production rates inside of stars, etc.~These values are rarely measured directly. For example, astronomical data sets consist almost exclusively of measurements of light. To connect this information to a model, one must employ knowledge of how light passes through a telescope, is recorded by electronics, and is corrected for a variety of calibration and bias effects. This knowledge is specialized, complex, and must be managed by anyone using the data, which is then multiplied by the number of instruments one seeks data from. The system we are proposing would internalize this knowledge, allowing a scientist to build a model and pass it into the framework \textit{while still described in terms of physical units}. The framework is also a large-scale store of public data which will be able to take the submitted model and apply it to all available data. Additionally, this system is designed to provide access to both the data and the models to other scientists, allow client applications for data visualization, as well as provide a platform for sharing models and education. We have given the moniker \textit{Deep Thought} \citep{hhgttg} to this framework.

\subsection{Interface to Deep Thought}

The aim of science is to understand nature; to do this we create models of natural phenomenon, devise experiments and make measurements, and analyze data to ask the question, ``How well does this data agree with my model?'' This process is at the heart of the proposed framework. We will explore four building blocks: data, models, computations, and libraries.

\subsubsection{Data}
Public data sets provide the foundation upon which the framework sits, where the aim is to collect as much data as possible into the system. Using astronomy as an example, a natural start would include the full imaging and spectral catalogs of SDSS, WISE, 2MASS, Tycho, Hipparcos, OGLE, Spitzer, Kepler, AKARI, etc.~All of the data would be made available through an API (see Section \ref{subsection:API}), and housed in such a way that any analysis may trivially access any other piece of data.

\subsubsection{Models}
The Model component provides an interface between astrophysical theory and observational data. Ultimately, any astrophysical model must make observable predictions in terms of images, spectra, or both. The framework accepts self-contained code from users (Models) which generate such physical predictions. The fundamental measure of goodness-of-fit for a Model is the probability of generating the observations from the model in question, or the likelihood. Using Models provided by researchers and full characterizations of survey properties, {\it Deep Thought} can calculate meaningful likelihoods for any catalog entry in the combined data set. If a provided model has free parameters, then these can be varied to find a best fit, or to characterize the parameter likelihood surface. 

For example: an astrophysicist hoping to model cool, dust-obscured stars might provide a Model that takes as parameters two temperatures and two luminosities. This model would generate a spectrum consisting of the sum of two blackbody curves. Uploading this Model to {\it Deep Thought}, the theorist asks it to find the best-fit temperatures for every object that appears as a point source in any or all of the SDSS, WISE, and 2MASS catalogs. Because the framework `knows' the point-spread function, photometric calibration, and other necessary properties for all of the surveys in its demesne, it can compute these likelihoods using all of the observations of each object, regardless of which surveys may have observed it. The best-fit parameters and their goodness-of-fit can be used to further winnow the point-source catalog for objects that are well fit, and with temperature parameters that fall in the range of interest to the investigator. The advantage here, what's really new, is that one can interrogate large data volumes using physical models throughout the process, without ever resorting to instrument quantities, confident that the translation is physically meaningful and has been performed by people who {\it are} intimately familiar with the instruments.

\subsubsection{Computation}

Some analyses require as an input an ensemble of data rather than a single value. For example, modeling the stellar luminosity function or a cross-correlation function of galaxies applied to large-scale structure is not performed on a per-source or image basis. To support these models, a module called a \textit{Computation} is introduced. Similar to a Model, it takes as input any ensemble of data and produces an output, which can be in the form of a function or data, which could be a single value, a new data set, or similar. A Computation could also produce an output value for a discrete set of objects. For example, an input could be a list of stars in a cluster, and the output could be distance, age, evolution stage for each member, etc.; such a calculation could not be made on a single star by star basis. The structure of the module would roughly match that of a Model; predefined inputs, outputs, documentation, etc. Model objects may have a Computation (or several) as a dependency, and vice versa.

\subsubsection{Libraries}

Some models are based on simulation or analytical generation of data. Modeling a stellar cluster requires a library of isochrones; modeling the spectrum of a stellar population or galaxy requires isochrone and spectral libraries which are produced through a mix of analytical models and observation. Large scale cosmological structure would need to be compared to mock catalogs. The \textit{Library} module is designed to contain such data that is either provided or not directly derived from archive data. Again, the module's structure would be similar to that of a Model and Computation, with the addition of a ``data''~directory (which is not to exclude a ``data''~directory from the other modules). Executable code may also play a part; an example might be a Library of the SFD dust map \citep{1998ApJ...500..525S} which would take an RA/dec input and return an extinction value, but require code and provided data to perform the calculation. Models and Computations may have dependencies on Libraries, and vice versa.

\subsection{Implementation}

\subsubsection{Data Storage}
The foundation of this framework is the requirement to a) store as much data as possible (or better, is available) together in b) such a way that it can be trivially cross-referenced. These are two separate, but related, problems. The storage of the data might be in a single location, or it may be distributed across several locations, as long as software services and high bandwidth makes it appear to be located in a single location. This can take the form of a server with sufficient disk space to hold the data sets required; today, even 200 TB is not prohibitively expensive, but it is beyond the resources of most astronomy departments. (Further, it is a waste of resources to duplicate such a system across a large number of institutions.) Another model currently exists in the form of commercial cloud platforms from companies such as Amazon, IBM, Microsoft, or Google. These solutions can easily manage the volume of data currently needed, with the advantage of offloading the management and maintenance responsibilities. Only the storage and bandwidth used would incur costs. Regardless, the simple storage of data is a commonly solved problem; further details are beyond the scope of this paper.

\subsubsection{Cross-Referencing Data Sets}
\label{section:cross-ref_data-sets}

The second component is the integration of data from multiple, very different, surveys. Merely putting data from different surveys on the same server does relatively little to allow a researcher to use them in concert. Correctly associating entries in different survey catalogs with one another is the next step, but even this can be problematic: for example, in crowded fields different PSFs or depths between surveys necessarily lead to ambiguous cross-association, and even when deblending is attempted no two survey pipelines use the same software with the same settings. Correctly comparing imaging or spectroscopic measurements from different instruments requires accounting for a host of issues, ranging from PSF and detector nonlinearities to differences in reduction and analysis pipelines. 

Fortunately, a more or less complete understanding of these technical issues is generally produced by individual surveys, and the process of generating an image or spectrum of a model, as it would have appeared in any given survey, is a fairly well-solved problem (see \citealt{2012AJ....144..188B} for an example of this approach). For example, a measurement of the color of a star in between WISE and SDSS survey bands might be difficult to interpret using an aperture magnitude from each survey, due to their very different PSFs, but relatively easy using the amplitudes of fitted models of the appropriate PSF from each survey. By collecting the technical survey information in one place, the framework will permit straightforward, simultaneous evaluation of models across all available survey data. The results of these will be queryable, allowing for meaningful comparison and selection of objects with observations taken by very different instruments. The same procedure permits easy use of multi-epoch observations.

Once this has been built, one can trivially make queries such as ``What are all of the available observations of galaxy \textit{X}?''~or~``What objects in the sky match a given set of parameters?'', without having to download any data. More importantly, one can query this data without even knowing what data is available; it's more natural to think in terms of the objects of research (e.g.~a location in the sky, a galaxy of interest, surface gravity) rather than having to manually extract this data from individual data sets where it's easily possible to overlook observations that might be available. It's preferable to move to a model where we automatically include all accessible observations that might be of interest as it's easier to exclude unwanted data than to seek out everything available.

\subsubsection{Application Programming Interface (API)}
\label{subsection:API}
The data contained in Deep Thought is clearly not intended (or in most cases possible) to be downloaded as a whole, either to a user's computer or even a single institution. A common workflow for scientists today is to locate data available online manually through a web form, download some subset in the form of flat files, and write code to read them into programs for analysis. If a new (or slightly different) data set is required, this process is repeated. One of the design goals, however, is to make data appear to the user as if it were local, which means providing a simple means of direct access regardless of how or where the data is actually stored. This can be accomplished through the publication of an open \textit{application programming interface} (API). An API is a source code-level description of how a program communicates with another piece of software or server. An example of such an interface is the SDSS API.\footnote{\url{http://api.sdss3.org}} A sample query to search for SDSS spectra within 300 arcseconds of the coordinates RA 200.14753\arcdeg~and declination 34.13938\arcdeg, between a redshift of 0.1 and 0.2, would look like this:

\begin{lstlisting}[breaklines=true,numbers=none]
http://api.sdss3.org/spectrumQuery?redshift=0.1-0.2&ra=200.14753d&dec=34.13938&radius=300
\end{lstlisting}

The query language would of course be generalized to the full domain of available data, published as a formal specification, and operate at the server level via HTTP. The advantage of using the same protocol in use by every web browser versus writing something new is that the rich variety of software and clients written for the world wide web can immediately be applied to retrieve scientific data. Rather than using (and learning) a different interface for each set of data (e.g.~a site- or data-specific web site), a consistent API across as many data sets as possible enables easy, convenient, and powerful data access. Further, using non-specialized protocols dramatically lowers the bar for anyone in the public to access the data and/or create their own interfaces without needing significant resources to store the data themselves.

Consider the difficulties (or at least tedium) of data access as described in Section \ref{section:workflows} above. Here, a query that describes a particular set of data can be written in code (while still being human-readable), such as ``all stars between 10 and 15 magnitudes within 3$^{\circ}$ of the Magellanic Clouds''~or~``a ten year history of GDP values of countries that have had the fastest population growth over the past year''. With an API, one could write a script containing the query, submit the query to Deep Thought, receive the results, then immediately start analyzing or visualizing the data. Listing \ref{listing:server_api} is an example how this might be done with only a few lines of code with no intermediate files or bookkeeping. Here, the query itself defines the subset of data desired, which has several advantages. First, the data does not need to be stored as a local file and managed; the query can simply be rerun. A query can be easily shared with a colleague who can then run it, saving the need to share files. Further, a query that is executed at a later date may reflect more, corrected, or updated data. Of course, saving data for other reasons (e.g.~reproducibility, performance) is certainly not excluded.

The API proposed in this framework would encompass both an interface to the repository of data (server API) and client interfaces that understand each type of data that is retrieved (client API). Complex data is necessarily stored in file formats that support structured data; HDF5, FITS, CDF, etc. This leads to a great deal of code to access the particular information sought, which requires specialized knowledge of multiple data formats and the specific organization for each data set. Often documentation is less than ideal. Analysis software typically follows the pattern of navigating a directory structure, locating a file, opening it, and finally reading the location in the file where the data is. This bookkeeping code is usually interspersed with the actual analysis code, making software difficult to read. Worse, roughly the same code is rewritten by virtually everyone who uses the data. A fully object-oriented, common API written for each file type (e.g.~a WISE image, an SDSS spectrum) would save time, be optimized by the community, and allow researchers to concentrate on analysis. Consider FITS files that contain SDSS spectra. If one were interested in getting the coordinates of the source object, one must go to the documentation to find which header data unit (HDU) the desired information is located, and write the specialized code to retrieve it. This is repeated for each datum required. Listing \ref{listing:client_api} shows an example of how the interface might look. In the first example, the code simply requires the file name of the data to initialize the object. Once defined, any metadata is immediately available from the object, regardless of the data format or the location of the data within the structure of the file. All of this logic is encoded into the implementation of the library; this feature alone can significantly reduce the amount of analysis code written. These objects will have rich relationships to other data sets. Here, a single method will associate the SDSS source to a matched source in the 2MASS catalog, from which any value can again be immediately accessed. In the second example, the spectrum object is created using the set of metadata that uniquely defines it. The code will understand the specifics of each data set; in the third example, an ``SDSSPlate''~object is defined from the ID number. Since a plate is used to measure a set of spectra, the library knows how to retrieve all of the spectra associated with that plate. More importantly, the same code would work whether the data is on the user's computer or not. The server-based API would be fully integrated into the file-level API; if the data is not locally available, it will connect to the repository, download it into a local cache (see Section \ref{section:caching_data}), and retrieve it.

\begin{lstlisting}[language=Python,keywordstyle=\color{blue},caption={Server API Example},label=listing:server_api]
import astropy.units as u
from astropy.coordinates import ICRS
from sdss.query import SpectrumQuery

query = SpectrumQuery()
query.redshift = [0.1, 0.2]
query.position = ICRS("09h55m52.7s +69d40m46s")
query.radius = 300 * u.arcsec

spectra = query.fetch()

for spectrum in spectra:
    print spectrum.ra, spectrum.dec
\end{lstlisting}

\begin{lstlisting}[language=Python,keywordstyle=\color{blue},caption={Client API Example},label=listing:client_api]
from surveys.sdss import SDSSPlate
from surveys.sdss.boss import BOSSSpectrum
from surveys.sdss.manga import MaNGADataCube

# access from file
s = BOSSSpectrum("spec-3982-55332-0483.fits")
print s.ra, s.dec, s.redshift
print s.emission_lines
print s.twomass_source.magnitude(band="j")

# access from other metadata
s = BOSSSpectrum(plate=6000, mjd=55555, fiber=12)
plot(s.wavelengths, s.flux)

p = SDSSPlate(id=4242)
for s in [x for x in p.spectra if x.type="sky"]:
    ... # analyze sky spectra
\end{lstlisting}

\subsubsection{Caching Data}
\label{section:caching_data}
A query executed via an API incurs some overhead for the remote system to retrieve the requested data and transmit it to the user. This may vary from a fraction of a second to potentially hours if the selection criteria is complex or requires significant computation. It would be against the aim of having data appear to be local if this overhead were experienced each time the query were executed (e.g.~a local script that is run multiple times). A multi-level caching scheme addresses this problem, and we can take particular advantage of the fact that most science archive data is read-only in nature.  The first time the data is retrieved, it would be downloaded to the user's computer (along with relevant metadata) into a local cache. Subsequent queries would check this cache first, then retrieve what is not present. A one terabyte drive on a desktop can be had at trivial cost and would provide an enormous cache for one user. Caching can also be performed at other levels, for example, a department (or even higher, a university) might provide many terabytes of space that would cache the most frequently accessed data. The caching algorithm would manage what data is stored locally, adapting over time to what is most used currently by its users. While potentially hundreds of terabytes might be available, most analyses involve selecting a considerably smaller subset for analysis. The API would assist in finding that data quickly, and caching will help with making it appear local (as, effectively, it will be, but without needing any management by the user). Finally, the data store itself might employ another level of caching to optimize performance. Further details are beyond the scope of this paper as these are well studied and solved problems in the commercial sector.

\subsubsection{Module Implementation}
\label{subsection:module_structure}

Model, Computation, and Library modules would be implemented as a package, defined as a directory containing files and folders that fully describe the module. A proposed scheme is presented here, where all elements are common to each type of module. Packages will contain a machine-readable file that lists dependencies on any external modules (e.g.~an ``AGB star''~Model might depend on a ``star''~Model), code library (e.g.~a Fourier transform library or Python packages), or specific system requirements (e.g.~Intel processor). Documentation of the implementation and usage of the module would be placed in a top-level directory, itself containing files in HTML, Markdown, and/or plain text (least preferred). This allows and encourages a rich display of the documentation which can contain images, rendered math equations, interactive JavaScript plots, live links to journal papers, etc. The mapping between the data input, the module, the parameters (if any), and the output should be explicitly defined in a data mapping file. An input might be a 2D array of pixel values, a spectrum, an entry in a point source catalog, etc. The number and type of parameters input, and structure of the data output (if any beyond a likelihood value) would also be explicitly defined.  To allow filtering of a large number of models, a keywords file would contain terms appropriate to the module. Another directory would contain the source (and optionally, executable) code, further organized by language. Any data that is provided as part of the model could be placed into a dedicated directory. If the data is too large, it should be placed into the data archive as a data set, and a file in the data directory would point to the set. This structure integrates very well with version control systems such as Git. Modules to be run can then be checked out of a Git repository by the system rather than submitted by hand, and it would be required that any module run be tagged with a specific version.

Once the data repository exists and the modules are defined, applying the models to the data is the next step. To first order, all available data that matches the defined inputs are passed to each model. For each Model (or more technically, unique combination of Model version number and input parameters) a new result data set is created. This set contains a unique identifier to the data input, the likelihood value for the Model, and any optional additional derived parameters. This result is of course associated with the model, but the inverse is also true; an investigation of the data will also return all of the models applied to it. Processing all available data through every model may seem unnecessary, but is in fact an important aspect of this framework. One important question that Deep Thought will be able to answer is, ``Which objects in our data are not sufficiently explained by any model?'' All Models must be applied to all data to ask this. We will still be able to achieve efficiencies through Model dependencies. For example, a Model looking for TP-AGB stars would start with the output of a Model that identified stars above a specified likelihood.

When researchers are developing a Model, they will have the option to apply it to a small part of the sky. Model development will be iterative; there's little point to expend the resources to apply a Model to the whole sky only to find a bug in the first few sources found. Once the user is confident with the model, it can then be applied to the full sky. In a similar manner, one could ``save'' part of the sky for a blind analysis.

\subsubsection{Costs}
A computational platform as described here will of course require funding to operate and maintain. The primary (non-personnel) costs are data storage, bandwidth, and CPU time. A first generation system can easily be hosted at a single institution. Funding could follow several models: grants to purchase bulk storage or a pool of CPU hours, a ``buy-in'' system where the cost is shared among member institutions, etc. Directed grants might award CPU time to historically black colleges and universities, minority universities, or universities in nations that are underrepresented in research.  An institution might contribute underutilized resources; for example, when a new Model is submitted, the actual computation could take place on several department servers that might otherwise be idle. The software might be sent to the department in the form of a pre-configured virtual machine which would be free to run at a low priority. Perhaps those who submit models might pay for their computation. If resources are sparse, the system might rely on distributed lazy loading. Imagine a desktop client application, where the user views a particular image. If a Model has not been applied to the objects in that image the user is interested in, they may allow their computer to do the computation immediately, where the results are returned to the system. Certainly there are other models that might be combined (such as employing cloud computing), and all of these would need to be considered to support a community-wide computation platform. This feature of the platform is one of its advantages; spread across the whole community, contributions would be affordable to all, and the large and small departments would have access to the same platform.

\section{Community Research \& Education}
\subsection{A First-Class Client Application}
We have presented above two of the three major components of the proposed framework: the computational platform and streamlined data access through APIs. The third major piece is an educational and social platform to connect these pieces to the user. Creating an API and simply expecting others to embrace it is not a reasonable approach; rather, a well designed application should also be designed alongside the machinery described above. This client would provide access to all the imaging, spectral, and catalog data available, and would be tightly integrated with all of the Models in the system. One would be able to search for Models by name, keywords, objects matched, or other criteria. When selected, the documentation that is built into the module would be displayed, again supporting rich interfaces such as HTML and JavaScript. Authors will be encouraged to effectively write a short article write as part of the module; the theory, the motivation, the means of calculation, etc.

While a ``social platform'' may seem trendy, it's worth noting that the world wide web was specifically invented for physicists to communicate. Journal publications represent the most formal means of scientific discourse, however, due to the rigor, formality, and frequency, the bar to communicate through that medium is very high. It is far easier to get someone to write a few hundred words (with diagrams) on their research than to write a journal paper. Further, due to their nature, most papers often require specialized knowledge or language to understand. We feel a more informal means to communicate research would be greatly beneficial, not only to researchers themselves but also students entering the field and even the public. (It's important to stress that this is a supplement to traditional publication, not a replacement.) Anyone would be able to contribute to the descriptions as well---effectively, this would be a Wikipedia for science research. There is an immense pool of knowledge that should be more easily disseminated. There are many scientists who have knowledge that should be put into textbooks but will never have the time to do so. Graduate students working on research projects are in the perfect phase to provide entries; they are still learning and can more easily write to those who are new as well, while having access to experts who can review their work without having to take the time to write the material themselves.

Some knowledge is built upon interactions with and observations from experts in the field. For example, if you were to sit down with the director of the SDSS survey and look at a few dozen spectra, you would hear, ``This is a good example of a star forming galaxy. This one shows there is a black hole in the center of galaxy; note the strength of the OIII line.'' If, through the use of such a client application, it was extremely easy to write such comments to be attached to individual spectra---or any data in the system---then anyone who comes across the same data will be able to learn from this expertise. Space is big; as it would be difficult to find random comments, one would be able to search for all objects that a particular person has commented on. This kind of interaction lowers the bar to the level of common social network tools, but would be a valuable communication and education tool for researchers. Members of the public who are interested in learning more about a particular subject would certainly find this useful as well.

\subsection{Enabling Technology}
The APIs discussed here would naturally be open to the public. There is considerable interest in science outside of the professional community, but access to and the interpretation of scientific data typically requires specialized knowledge. Even a service that serves astronomical images as tiles can be used to build an interface by someone with no formal astronomy background. Open APIs would significantly lower the bar to access the data, enabling anyone to build applications and visualizations while not requiring significant computing or storage requirements. Individual surveys can create desktop or bespoke web applications that highlight a particular set of data or research project or provide custom interfaces to complex data. The city of New York, for example, has had success with people creating visualizations after making municipal data available to the public.\footnote{\url{https://data.ny.gov}} Providing APIs to the public will also enable a new class of applications that run on mobile devices with constrained resources (smartphones, tablets). There is great potential in the exploration of interfaces for scientific analysis and visualization in touch devices, particularly as the screen resolution and processing power have dramatically increased over the past few years. For example, such a program on a tablet with access to the full imagining catalogs of WISE, SDSS, and more, would enable anyone to study data, alleviating the need to write code to retrieve data, to store it, to search through hierarchies of files on disk, etc.~It may be that providing such access to data in this way may be one of the biggest (and unpredictable) enablers of this entire proposal.

\subsection{Proprietary Data}
Researchers are used to performing their own analyses and keeping the results private until publication. Often, observational data has some proprietary period. It would be counterproductive to restrict this platform to data that is not (yet) publicly available, but this must be balanced with the open and collaborative nature of science. One solution may be to allow the addition of closed data sets which have limited access, but only with the condition that the data be made public after a specified period. Models can be contributed under the same condition; they may be run and the results saved, but remain closed for an accepted period of time. The authors favor a policy of use for the system that any Model used in a paper be cited and made publicly available upon publication in the spirit of transparent, repeatable science.

\subsection{Existing Services}
It is worth spending a few words comparing the framework here to a few existing services. The Sloan Digital Sky Survey has a web interface that serves SDSS data covering imaging and spectral data. The SkyServer provides an image search function, while the CasJobs service allows database queries to be submitted as background jobs. The WISE and 2MASS point source catalogs are included in the database for cross referencing with SDSS data, but adding new data sets of more than a several thousand entries is not easy, and beyond that, it is impossible. These services are good for working with small to medium sized data sets (from hundreds to thousands of records), but it's easy to quickly run into the limits of a web interface and the per-user quotas. The Virtual Observatory is broader in scope than the SkyServer and aims to provide access to a larger volume of astronomical data, spread across multiple institutions.

The emphasis of both these services, though, is to primarily provide data; neither are computing platforms, and it is expected that researchers download the data they need to integrate into their own analysis. (This is of course not to take away from these and similar services; they are not designed to provide a computing platform, and often are intended to simply provide access to data.) As outlined above, this cannot scale; if one were interested in testing a particular model against every point source in the WISE catalog while integrating observations from 2MASS, SDSS, and a user-generated mock catalog, for example, this is far too much data to download to a single machine. Even if all of this data were downloaded, it would be in the form of flat files, leaving the researcher with the significant task of organizing it into a database, creating the cross references, optimizing searches, etc. This is to say nothing about comparing dissimilar photometric apertures, point spread functions, etc.~(see Section \ref{section:cross-ref_data-sets}). A great deal of specialized computer science knowledge must be applied before science can even begin (multiplied by the number of scientists using the data). Another important point is that searches using any of these services can only return binary results; either an item in the catalog satisfies the query arguments exactly or it does not. For example, the SDSS photometric database identifies whether an entry is a star or a galaxy (or something else). This identification is based on a model, but is not 100\% correct. Using the platform presented here, anyone could create their own model, generate a catalog that anyone could use, and ask the question, ``Which objects have not been categorized to a better than 90\% likelihood?'' It is the edge cases that are often the most interesting to explore, the ones that challenge our models, that we aim to make easy to search for.  By offloading the storage and bookkeeping requirements and proving researchers with a computing platform that is centered around models rather than tables with rows of numbers, researchers will be able to handle far greater volumes of data than they can presently while shifting the focus back to research and exploration of data where it belongs.

\section{Conclusions}
We have presented here a framework designed to enable a new way to do science, taking advantage of the large volumes of data now available (and to come) in a way that hasn't been possible before, going beyond existing interfaces to data. The core concepts are the creation and integration of an astronomical data repository, the cross referencing of this data that understands and incorporates instrumental effects, transparent and simple programmatic access to the data, the ability to apply an arbitrary theoretical model on a computing platform, and an integrated education tool for researchers and the public alike. Managing access to increasingly larger data sets is necessary to enable us to ask the important questions. Do we understand the data? How accurate are our models? Do our models hold up with new data? What is in our data that does not fit any of our models? This last question is one of the most compelling that we look forward to addressing with this framework.



\bibliographystyle{apj}
\bibliography{data_science}


\end{document}